\documentclass[prd,twocolumn,nofootinbib,aps,superscriptaddress]{revtex4}
\usepackage{latexsym}
\usepackage{graphicx}
\usepackage{caption2}
\usepackage{psfrag}
\usepackage{amsmath}
\usepackage[hypertex]{hyperref}

\usepackage{amssymb} 
\usepackage{epsfig,psfrag}
\usepackage{longtable}

\input epsf

\def\nslash{n\!\!\!\slash}
\def\bnslash{{\bar{n}}\!\!\!\slash}

\def\Dslash{D\!\!\!\!\slash}

\def\pslash{p\!\!\!\slash}

\newcommand{\bea}{\begin{eqnarray}}
\newcommand{\eea}{\end{eqnarray}}
\newcommand{\nn}{\nonumber\\}
\newcommand{\dslash}[1]{#1 \mspace{-10mu}/\mspace{3mu}}

\def\cD{{\cal D}}
\def\cB{{\cal B}}
\def\cDslash{{\cal D}\!\!\!\!\slash}

\def\nn{\nonumber}

\newcommand{\bn}{{\bar{n}}}
\newcommand{\mcdot}{\!\cdot\!}

\begin{document}

\title{On different ways to quantize Soft-Collinear Effective Theory} 
\author{Christian W. Bauer}

\affiliation{Ernest Orlando Lawrence Berkeley National Laboratory, University of California, Berkeley, CA 94720}
\author{Oscar Cat\`a}

\affiliation{Ernest Orlando Lawrence Berkeley National Laboratory, University of California, Berkeley, CA 94720}
\author{Grigory Ovanesyan}

\affiliation{Ernest Orlando Lawrence Berkeley National Laboratory, University of California, Berkeley, CA 94720}
\begin{abstract}
Collinear fields in soft collinear effective theory (SCET) can be made invariant under collinear gauge transformations by multiplying them with collinear Wilson lines. We discuss how we can quantize SCET directly in terms of these gauge invariant fields, allowing to directly calculate $S$ matrix elements using the gauge invariant collinear fields. We also show how for each collinear direction SCET can be written in terms of fields whose interactions are given by the usual QCD Lagrangian, and how external operators coupling these different directions can be constructed. 
\end{abstract}
\maketitle

Soft collinear effective theory (SCET) \cite{Bauer:2000ew, Bauer:2000yr,Bauer:2001ct,Bauer:2001yt} is by now a rather mature effective field theory with wide applications in $B$ physics and collider physics. SCET describes QCD in the kinematic regime where the energy of particles is far in excess of their (invariant) mass. Short distance physics is contained in Wilson coefficients which are determined order by order in perturbation theory. Long distance physics on the other hand is described by separate collinear fields for each light-like direction, together with Wilson lines $Y_n$ describing the usoft physics interactions between the different collinear directions. Since there are no direct interactions between collinear fields in different directions, gauge invariance requires the presence of Wilson lines to render collinear fermions and gauge bosons gauge invariant. 

\begin{figure}[t!]
\begin{align}
V^{(1)}_{{\cal L}_I} &= & 
\begin{picture}(30,10)(11,20)
     \mbox{\epsfxsize=1.5truecm \hbox{\epsfbox{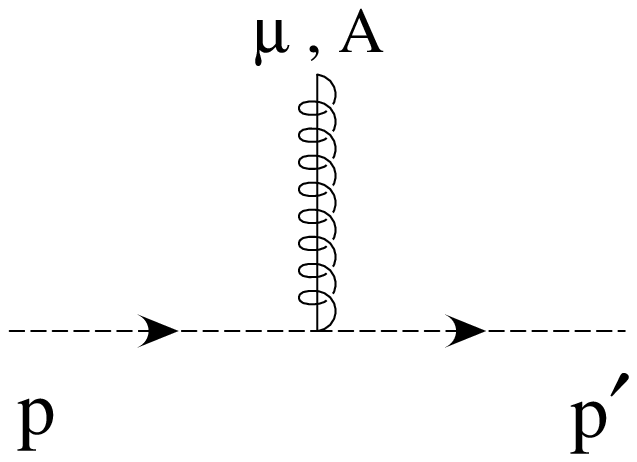}}  }
  \end{picture}
  &
  \hbox{$  = 
  ig T^A\!\left[n^{\mu}+\frac{\gamma^{\mu}_{\perp}\pslash_{\perp}}{\bar{n}\cdot p}+\frac{\pslash'_{\perp}\gamma^{\mu}_{\perp}}{\bar{n}\cdot p'}-\frac{\pslash_{\perp}'\pslash_{\perp}}{\bar{n}\cdot p' \bar{n}\cdot p}\bar{n}^{\mu}\right]\!\frac{\bnslash}{2}
  $ } 
  \nn\\\nn\\[-5pt]
V^{(2)}_{{\cal L}_I} &= & 
\begin{picture}(30,10)(11,20)
     \mbox{\epsfxsize=1.5truecm \hbox{\epsfbox{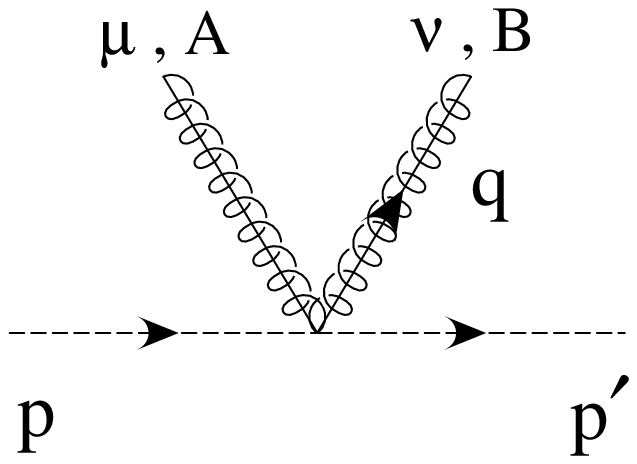}}  }
  \end{picture}
  &
  \hbox{$  = ig^2 \!\left[
  \frac{T^A T^B}{\bar{n}\cdot(p-q)}\gamma_{\perp}^{\mu}\gamma_{\perp}^{\nu}
  +\frac{T^B T^A}{\bar{n}\cdot(q+p')}\gamma_{\perp}^{\nu}\gamma_{\perp}^{\mu}\right]\!\frac{\bnslash}{2} + (\ldots)
 $ }
  \nn\\\nn\\[-5pt]
V^{(1)}_{{\cal L}_{II}} &= & 
\begin{picture}(30,10)(11,20)
     \mbox{\epsfxsize=1.5truecm \hbox{\epsfbox{FR1.eps}}  }
  \end{picture}
  &
  \hbox{$  = 
V^{(1)}_{{\cal L}_{I}} +i g T^A\!\left[\frac{1}{\bar{n}\cdot(p-p')}\left(\frac{p^2}{\bar{n}\cdot p}-\frac{p'^2}{\bar{n}\cdot p'}\right)\bar{n}^{\mu}\right]\!\frac{\bnslash}{2}
 $ }
  \nn\\\nn\\[-5pt]
V^{(2)}_{{\cal L}_{II}} &= & 
\begin{picture}(30,10)(11,20)
     \mbox{\epsfxsize=1.5truecm \hbox{\epsfbox{FR2.eps}}  }
  \end{picture}
  &
  \hbox{$  = 
V^{(2)}_{{\cal L}_{I}}  + (\ldots)
 $ }
  \nn\\\nn\\[-5pt]
V^{(1)}_{{\cal L}_{III}} &= & 
\begin{picture}(30,10)(11,20)
     \mbox{\epsfxsize=1.5truecm \hbox{\epsfbox{FR1.eps}}  }
  \end{picture}
  &
  \hbox{$  = 
  ig T^A\!\left[n^{\mu}+\frac{\gamma^{\mu}_{\perp}\pslash_{\perp}}{\bar{n}\cdot p}+\frac{\pslash'_{\perp}\gamma^{\mu}_{\perp}}{\bar{n}\cdot p'}\right]\!\frac{\bnslash}{2}
  $ } 
  \nn\\\nn\\[-5pt]
V^{(2)}_{{\cal L}_{III}} &= & 
\begin{picture}(30,10)(11,20)
     \mbox{\epsfxsize=1.5truecm \hbox{\epsfbox{FR2.eps}}  }
  \end{picture}
  &
  \hbox{$  = ig^2 \!\left[
  \frac{T^A T^B}{\bar{n}\cdot(p-q)}\gamma_{\perp}^{\mu}\gamma_{\perp}^{\nu}
  +\frac{T^B T^A}{\bar{n}\cdot(q+p')}\gamma_{\perp}^{\nu}\gamma_{\perp}^{\mu}\right]\!\frac{\bnslash}{2}
 $ }
  \nn\\\nn\\[-5pt]
\Delta_{{\cal L}_{III}} &= & 
\begin{picture}(30,10)(11,0)
     \mbox{\epsfxsize=1.5truecm \hbox{\epsfbox{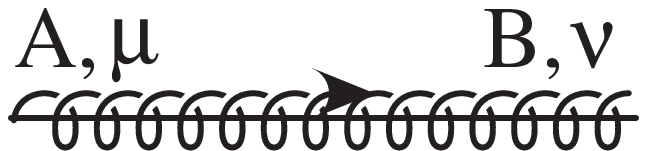}}  }
  \end{picture}
  &
  \hbox{$  = 
-i \frac{\delta^{AB}}{k^2+i\epsilon}\left(g_{\mu\nu}-\frac{{\bar{n}}_{\mu}k_{\nu}+{\bar{n}}_{\nu}k_{\mu}}{{\bar{n}}\cdot k}\right) $ }
\nn\\[-15pt]\nn
\end{align}
{\caption[1]{A subset of Feynman rules for the three different formulations of SCET. The (...) denote terms  which do not contribute to the tadpole diagram of Fig.~{\ref{fig1}}b) in Feynman gauge.}
\label{feynmanrules} 
}
\end{figure}

There are several equivalent versions of SCET used in the literature. The original formulation of SCET~\cite{Bauer:2000ew, Bauer:2000yr} described the interactions between the gauge dependent collinear quark $\xi_n$ and gluon $A_n$ fields, with the leading order Lagrangian given by\footnote{We here omit any reference to gauge-fixing and potential ghost terms. We will address the quantization of gauge fields later on.}
\begin{eqnarray}
{\cal{L}}_{I}^n(\xi_n,A_n)&=&{\bar{\xi}}_n\left[in\mcdot D_n+i\Dslash_n^{\perp}\frac{1}{i{\bn}\mcdot D_n}i\Dslash_n^{\perp}\right]\frac{\bnslash}{2}\xi_n\nn\\
&& - \frac{1}{2} {\mathrm{Tr}}\, F^{n}_{\mu\nu} F_n^{\mu\nu}\, ,\label{L1}
\end{eqnarray}
with the standard definition of the covariant derivative and the field strength tensor
\begin{equation}
iD_n^{\mu}=i\partial^{\mu}_n+g_sA_n^{\mu} \,, \qquad F_n^{\mu\nu} = \frac{i}{g_s} [ D_n^{\mu},  D_n^{\nu}]\,,\\
\end{equation}
where the partial derivative $\partial_n$ is given in terms of the label operator introduced in~\cite{Bauer:2001ct}
\begin{equation}
i\partial_n^\mu = \bn \mcdot {\cal P} \frac{n^\mu}{2} + {\cal P}_\perp^\mu + i n \mcdot \partial \frac{\bn^\mu}{2}\,.
\end{equation}
In order to construct gauge invariant operators containing collinear fermions, these fermions are required to appear in the gauge invariant combination
\begin{equation}
\label{chidef}
\chi_n=W_n^{\dagger} \xi_n\,,\\
\end{equation}
where $W_n$ is the collinear Wilson line \cite{Bauer:2001ct}
\begin{equation}
W_n = {\rm P} \exp \left[ - i g_s \! \int_0^\infty \! {\rm d} s \,\,\bn \mcdot A_n(\bn s + x) \right]\,.
\end{equation} 
Using a simple field redefinition, one can easily obtain the collinear Lagrangian in terms of these gauge invariant combinations
\begin{eqnarray}
{\cal{L}}_{II}^n(\chi_n,A_n)&=&{\bar{\chi}}_n W^{\dagger}_n\left[in\mcdot D_n+i\Dslash_n^{\perp}\frac{1}{i{\bn}\mcdot D_n}i\Dslash_n^{\perp}\right]\frac{\bnslash}{2}W_n\chi_n\nn\\
&& - \frac{1}{2} {\mathrm{Tr}}\, F^{n}_{\mu\nu} F_n^{\mu\nu}\,.\label{L2}
\end{eqnarray}
Since the fields $\chi_n$ are gauge invariant, the combination $W_n^\dagger D_n^\mu W_n$ has to be gauge invariant as well. Thus, we can define \cite{Arnesen}
\begin{equation}
{\cal{D}}_n^{\mu}=W_n^{\dagger}D_n^{\mu}W_n\,.
\end{equation}
The gauge invariant derivative operator ${\cal{D}}_n^{\mu}$ can be written in terms of the partial derivative and a gauge invariant gluon field ${\cal{B}}_n^{\mu}$
\begin{equation}
i{\cal{D}}_n^{\mu}=i\partial_n^{\mu}+g_s{\cal{B}}_n^{\mu}\,,
\end{equation}
where
\begin{equation}
\label{Bdef}
{\cal{B}}_n^{\mu} = \left[ \frac{1}{\bn \mcdot \partial} \left[ i \bn \mcdot \cD_n , i \cD_n^\mu \right ]\right] = \frac{1}{g_s}\left[W_n^\dagger iD_n^\mu W_n\right]\,,
\end{equation}
and the derivatives only act within the square brackets. In terms of these fields, the Lagrangian reads
\begin{eqnarray}
{\cal{L}}_{III}^n(\chi_n,{\cal B}_n)&=&{\bar{\chi}}_n\left[in\mcdot \cD_n+i\cDslash_n^{\perp}\frac{1}{i{\bar{n}\mcdot\partial}}i\cDslash_n^{\perp}\right]\frac{\bnslash}{2}\chi_n\nn\\
&& - \frac{1}{2} {\mathrm{Tr}}\,{\cal{F}}^{n}_{\mu\nu} {\cal{F}}_n^{\mu\nu}\,,\label{L3}
\end{eqnarray}
where we have defined
\begin{equation}
\label{curlyFdef}
{\cal F}_n^{\mu\nu} = \frac{i}{g_s} [ {\cal{D}}_n^{\mu},  {\cal{D}}_n^{\nu}]\,.
\end{equation}
A sample of Feyman rules for the three different formulations of SCET is shown in Fig.~\ref{feynmanrules}.

It is well known that the dynamics of SCET with a single collinear direction is identical to full QCD. This is of course expected, since one can perform a simple Lorentz boost along the direction $n$ to make all momentum components of the collinear field similar in magnitude. Since this eliminates any large ratio of scales, the interactions have to be those of full QCD. This implies that for example the wave function renormalization in SCET is equivalent to that of full QCD, as was first shown in~\cite{Bauer:2000ew,Bauer:2000yr}. This equivalence has been used in the literature in order to simplify perturbative calculations in SCET ({\it{vid.}}, for instance, \cite{Becher:2006qw}).  

It is the purpose of this paper to study the relationship between different formulations of SCET. We work out the relationship between SCET using gauge dependent and gauge invariant degrees of freedom, as well as the relationship between full QCD and collinear fields in a single direction further. One of the features of the original formulation of SCET is that collinear gluons are coupled to the quark fields in a non-linear way. This means that there are an infinite number of vertices consisting of quark-antiquark and an arbitrary number of collinear gluons, whose Feynman rules get increasingly complicated. This makes the theory particularly unfriendly for computations beyond the one-loop order. 

We will show how to quantize SCET directly in terms of the gauge invariant degrees of freedom, and write the theory as a path integral over these gauge invariant fields. We will also discuss how to re-express the theory using only the interactions of full QCD. This first gives a precise field theoretical understanding of the well known property of SCET that the dynamics in a given collinear direction are equivalent to that of full QCD. Our formulation using directly the generating functional will extend this result to include interactions between different collinear directions through local operators. One can hope that these results will simplify the perturbative calculation of matching coefficients in the future, since much of the SCET calculations are now identical to the corresponding QCD results. 

\begin{figure}[!t]
\centerline{\mbox{\epsfysize=0.95truecm \hbox{\epsfbox{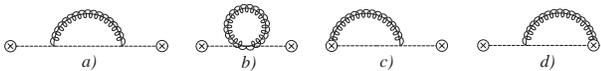}}  }}
{\tighten \caption[1]{Diagrams contributing to the gauge invariant jet
function at one loop.}
\label{fig1} }
\end{figure}

It will prove instructive, however, to first illustrate this equivalence between different formulations of SCET using a simple one-loop calculation. Consider the two point correlator of two gauge invariant fermion fields
\begin{equation}
\langle0|T \chi_n(x) \bar{\chi}_n (y) |0\rangle = \langle0|T W^\dagger_n(x) \xi_n(x) \bar{\xi}_n (y) W_n(y) |0\rangle\,.
\end{equation}
The Fourier transform of this correlator is what is known in the literature as the jet function, and plays a crucial role in any process containing external collinear particles. In the original formulation of SCET in terms of $\xi_n$ and $A_n$ fields there are four diagrams contributing at one loop, which are shown in Fig.~\ref{fig1}. The first two diagrams are entirely built out of interactions contained in the Lagrangian of the theory, while in the last two diagrams one of the gluon couplings comes from the Wilson lines $W_n$ or $W_n^\dagger$. Using the Feynman rules given in Fig.~\ref{feynmanrules}, one can easily obtain the result
\begin{widetext}
\begin{align}
  &D_{I,a}=g_s^2 \frac{\dslash{n}}{2} \frac{\bn\mcdot p}{p^2} C_F (2-D)\int\frac{d^{D}k}{(2\pi)^D}\left[\frac{1}{2}\frac{1}{(k^2+i\epsilon)((k+p)^2+i\epsilon)}-\frac{\bn \mcdot p}{p^2}\frac{1}{(k^2+i\epsilon)\bn\mcdot (k+p)}\right]\label{DIA}\,,\\
  &D_{I,b}=g_s^2 \frac{\dslash{n}}{2} \left(\frac{\bn \mcdot p}{p^2}\right)^2 C_F (2-D)\int\frac{d^D k}{(2\pi)^D}\frac{1}{(k^2+i\epsilon)\bn\mcdot (k+p)}\label{DIB}\,,\\
&D_{I,c}=D_{I,d}=g_s^2 \frac{\dslash{n}}{2} \frac{\bn\mcdot p}{p^2} C_F( n\mcdot \bn) \int\frac{d^D k}{(2\pi)^D}\frac{\bn\mcdot (k+p)}{(k^2+i\epsilon)((k+p)^2+i\epsilon)\bn\mcdot k}\,.
\end{align}
\end{widetext}
Note that the tadpole diagram is canceled exactly against the second term in the first diagram. Performing the remaining integrals and summing the diagrams one obtains the well known result~\cite{Manohar:2003vb}
\begin{align}
&D_I=i\frac{\alpha_s C_F}{4\pi}\frac{\dslash{n}}{2} \frac{\bn\mcdot p}{p^2}\left(\frac{\mu^2}{-p^2}\right)^\epsilon \left[\frac{4}{\epsilon^2}+\frac{3}{\epsilon}+7-\frac{\pi^2}{3}\right]\,.
\end{align}

We can repeat this calculation using the formulation of SCET in terms of $\chi_n$ and $A_n$ fields. This removes the last two diagrams of Fig.~\ref{fig1}, since there are no Wilson lines in the definition of the correlator when written in terms of $\chi_n$ fields. However, the extra Wilson lines in the collinear Lagrangian change the Feynman rules in the way shown in Fig.~\ref{feynmanrules}. While this does not change the result for the second diagram, the first diagram is now
\begin{widetext}
\begin{eqnarray}
D_{II,a} & = &g_s^2 \frac{\dslash{n}}{2} \frac{\bn \mcdot p}{p^2} C_F \int\frac{d^{D}k}{(2\pi)^D}\Bigg[(2-D)\left(\frac{1}{2}\frac{1}{(k^2+i\epsilon)((k+p)^2+i\epsilon)}-\frac{\bn \mcdot p}{p^2}\frac{1}{(k^2+i\epsilon)\bn \mcdot (k+p)}\right)\nonumber\\
 && +  2n\mcdot \bn\frac{\bn \mcdot (k+p)}{(k^2+i\epsilon)((k+p)^2+i\epsilon)\bn \mcdot k}-(n\mcdot \bn)^2\frac{\bn \mcdot p}{p^2}\frac{1}{(k^2+i\epsilon)\bn \mcdot k}\Bigg]\,.
\end{eqnarray}
\end{widetext}
The first two terms reproduce the result for $D_{I,a}$, and the third reproduces $D_{I,c}+D_{I,d}$. Finally, the fourth term in $D_{II,a}$ vanishes, since it is odd as $k \to -k$. Thus, the sum of all diagrams is identical in both versions of the theory. 

Showing that we can reproduce this result using the fully gauge invariant $\chi_n$ and $\cB_n$ fields is a little more tricky. This is because now there are no Wilson lines whatsoever, neither in the definition of the correlator nor in the Lagrangian of Eq.~(\ref{L3}). Thus, it is not immediately obvious how the contributions from diagrams $D_{c}$ and $D_{d}$ are reproduced in this case. However, care has to be taken when deriving the gluon propagator. The $\cB_n$ field is by construction explicitly invariant under collinear gauge transformations, and the usual procedure of adding an arbitrary gauge-fixing term to the Lagrangian is not valid. However, from the definition of $\cB_n^\mu$ in Eq.~(\ref{Bdef}) one easily verifies the constraint $\bn \mcdot \cB_n = 0$ (see Eqs.~(\ref{constraint}) and (\ref{parallel}) below). Thus, the propagator of a $\cB_n$ field has to satisfy the condition $\bn_\mu \Delta_\cB^{\mu\nu} = 0$. As we will discuss in more detail later, the propagator takes the form \cite{Bassetto:1984dq, Veliev:2001gp}
\begin{equation}\label{propB}
(\Delta_{\cB})_{ \mu\nu}^{ab}(k)=\frac{-i\delta^{ab}}{k^2+i\epsilon}\left(g_{\mu\nu}-\frac{{\bar{n}}_{\mu}k_{\nu}+{\bar{n}}_{\nu}k_{\mu}}{{\bar{n}}\mcdot k}\right)\,.
\end{equation}  
Using this propagator for the gauge invariant gluon field one can easily verify that
\begin{equation}
D_{III,a} =  D_{II,a}\,, \quad D_{III,b} =  D_{II,b}\,.
\end{equation}
Thus, the three formulations of SCET give identical results to one another for the two point correlator of two gauge invariant collinear fermion fields. 

In order to generalize this discussion to any matrix element, we quantize the theory directly in terms of the various fields. This is achieved by using the path integral formulation, working directly with the generating functional of the theory
\begin{equation}
Z[J]=\int\! {\cal{D}}{\bar{\xi}}_n{\cal{D}}\xi_n{\cal{D}}A_n^{\mu}\,{\mathrm{exp}}\left[i\int \! d^4x\, {{\cal{S}}_I}(\xi_n,A_n^{\mu},J_n)\right]\,,
\label{Zdef}
\end{equation}
where we have defined
\begin{eqnarray}
\label{actionSI}
{\cal{S}}_I&=&\sum_n\big[{\cal{L}}^n_I+{\bar{J}}_n^\xi \xi_n+{\bar{\xi}}_n J_n^\xi+{\bar{J}}_n^\chi W^\dagger_n\xi_n+{\bar{\xi}}_nW_n J_n^\chi\nn\\
&&+J_{n\mu}^A A_n^\mu+J_{n\mu}^{\cal B}{\cal B}_n^\mu(A_n)\nn\\
&&+\sum_k {J}_{k}{\cal O}_k\left(W_n^\dagger\xi_n,{\cal B}_n^\mu(A_n)\right)\,.
\end{eqnarray}
A few comments are in order to understand our notation. First, the integration in Eq.~(\ref{Zdef}) is over all fields with different directions $n$. Second, the subscripts $I,II,III$ indicate which version of SCET we are using, with Lagrangians given in Eqs.~(\ref{L1}), (\ref{L2}) and (\ref{L3}) above. Third, we have added separate currents for the gauge invariant fields $\chi_n = W_n^\dagger \xi_n$ and $\cB_n^{\mu} = \cB_n^{\mu}(A_n) = \frac{1}{g_s} [ W_n^\dagger iD_n^{\mu} W_n ]$, as well as for the gauge dependent fields $\xi_n$ and $A_n$. This allows us to calculate correlators with gauge invariant fields, such as the jet function, as well as those with gauge dependent fields, as is often done in matching calculations to QCD. Finally, we have indicated currents $J_k$ for any local operator in SCET. Such operators are typically written in terms of the gauge invariant fields, and an example would be the production current for two collinear fields in opposite directions, ${\cal O}_2 = \bar \chi_{n} \Gamma \chi_{\bn}$. 

In order to obtain the generating functional with the Lagrangian written in terms of $\chi_n$ fields, we make the field redefinition given in Eq.~(\ref{chidef}), which just amounts to a change in the integration variable in the generating functional. Since $W_n^\dagger W_n = 1$, one can easily show that the integration measure is the same when written in terms of the $\chi_n$ fields
\begin{equation}
   {\cal D} \xi_n {\cal D}\bar{\xi}_n {\cal D} A_n^{\mu}= {\cal D}\chi_n {\cal D}\bar{\chi}_n  {\cal D} A_n^{\mu}\,.
   \label{measureOneTwoo}
\end{equation}
Thus, the generating functional can be written as
\begin{equation}
Z[J]=\int \!{\cal{D}}{\bar{\chi}}_n{\cal{D}}\chi_n{\cal{D}}A_n^{\mu}\,{\mathrm{exp}}\left[i\!\int \! d^4x\, {\cal{S}}_{II}(\chi_n,A_n^{\mu},J_n)\right]\,,\end{equation}
with 
\begin{eqnarray}
{\cal{S}}_{II}&=&\sum_n\big[{\cal{L}}^n_{II}+{\bar{J}}_n^\xi W_n\chi_n+{\bar{\chi}}_nW_n^\dagger J_n^\xi+{\bar{J}}_n^\chi \chi_n+{\bar{\chi}}_n J_n^\chi\nn\\
&&+J_{n\mu}^A A_n^\mu+J_{n\mu}^{\cal B} {\cal B}_n^\mu(A_n)\big]
\nn\\
&&+\sum_k {J}_{k}{\cal O}_k(\chi_n,\cB_n(A_n))\,.
\end{eqnarray}
In other words, any matrix element written in terms of $\xi_n$ and $A_n$ fields is identical to the matrix element written in terms of $\chi_n$ and $A_n$ fields, as long as the interactions between the fields are given by the Lagrangian ${\cal L}_{II}$ instead of ${\cal L}_{I}$. 

Next, we discuss the relation between the gauge dependent gluon field $A_n$ and the gauge invariant field ${\cal B}_n$. The Yang Mills action is given by
\begin{equation}
\label{ZYM}
Z_{\rm YM} =\int \!{\cal D}A_n^\mu \, e^{iS_{\rm YM}[A_n]} \,,
\end{equation}
where
\begin{equation}
\label{SYM_A}
S_{\rm YM}[A] = -\frac{1}{2} \int \! d^4 x \, \sum_n {\rm Tr}\,F^n_{\mu\nu}F_n^{\mu\nu}\,.
\end{equation}
Recall that the relation between these two fields is given by
\begin{equation}
\label{ABrelation}
{\cal{B}}_n^{\mu}=\frac{1}{g_s}\left[W_n^{\dagger} iD_n^\mu  W_n\right]\,,
\end{equation}
where $\partial_n^{\mu}$ acts only within the square brakets. Since the Wilson lines $W_n$ are unitary, the Yang-Mills action can be written in terms of the ${\cal B}_n^{\mu}$ fields as
\begin{equation}
\label{SYM_B}
S_{\rm YM}[{\cal B}] = -\frac{1}{2} \int \! d^4 x \, \sum_n{\rm Tr}\, {\cal F}^n_{\mu\nu}{\cal F}_n^{\mu\nu}\,,
\end{equation}
where ${\cal F}_n^{\mu\nu}$ is given in Eq.~(\ref{curlyFdef}). 
However, in order to write the generating functional in terms of the fields ${\cal B}_n$ requires changing the integration measure as well, and that is where additional care has to be taken. From the definition of the ${\cal B}_n$ field we can immediately see that
\begin{equation}\label{constraint}
\bn \mcdot {\cal B}_n = 0\,,
\end{equation}
which follows from the well known relation of Wilson lines
\begin{equation}\label{parallel}
\bn \mcdot D_n \, W_n = W_n \,\bn \mcdot \partial_n \,.
\end{equation}
Thus, while there are four components of the $A_n^\mu$ field, there are only three components for the ${\cal B}_n^{\mu}$ field, making the Jacobian for the change in the integration measure singular. 

Of course, the fact that the ${\cal B}_n^\mu$ field has less independent components than the $A_n^\mu$ field is not unexpected, given that the former is gauge independent, while the latter contains all the gauge redundancy. The only way one can obtain a meaningful definition of a Jacobian factor is by removing the gauge redundancy and thus considering only three of the four components of the $A_n^\mu$ field. This can be achieved using the usual Faddeev-Popov procedure, by inserting a representation of unity into the path integral (\ref{ZYM}) in the following form
\begin{equation}
\label{FPunity}
1 = \int \! {\cal D}\alpha(x) \, \delta[G(A_n^\alpha)] \, \det\left(\frac{\delta G(A_n^\alpha)}{\delta \alpha} \right)\,,
\end{equation}
where $G(A_n^\alpha)$ is some gauge-fixing function linear in the gauge field. Here $\alpha(x)$ defines a specific gauge transformation and $A_n^\alpha$ denotes the (infinitesimally) gauge transformed field
\begin{equation}
(A^{\mu}_n)^\alpha = A^{\mu}_n + \frac{1}{g_s}D^{\mu}_n\alpha\,.
\end{equation}
Note that for infinitesimal gauge transformations (from which all finite transformations can be constructed),  the determinant of $\delta G / \delta \alpha$ is in general a function of $(A^{\mu}_n)^\alpha$ but independent of $\alpha$.

Following the standard treatment, the gauge invariance of both the action and the integration measure allows one to write
\begin{equation}
Z_{\rm YM} = \int \!{\cal D}\alpha \int \!{\cal D}A_n^\mu \, \delta[G(A_n)]\,E_G[A_n]\,,
\end{equation}
where we have defined
\begin{equation}
E_G[A_n] = \det \left(\frac{\delta G(A_n^\alpha)}{\delta \alpha} \right)\!\![A_n]\, e^{iS_{\rm YM}[A_n]}\,.
\end{equation}
One should remember that the determinant ${\rm det}(\delta G/  \delta \alpha)$ is independent of $\alpha$ and therefore the integral over the gauge freedom is just a global factor that can be safely ignored. The important feature of this way of writing the path integral is that the integration measure ${\cal D} A_n^\mu \,\delta[G(A_n)]$ contains only three components of the $A_n^\mu$ field, and can thus be related to the integration measure of the ${\cal B}_n^\mu$ field. This allows us to formally write
\begin{equation}
{\cal D}A_n^{\mu} \, \delta[G(A_n)] = J_G[{\cal B}_n] \, {\cal D}{\cal B}_n^{\mu} \, \delta[\bn \mcdot {\cal B}_n] \,,
\end{equation}
where the Jacobian factor for the change of the integration measure $J_G[{\cal B}_n]$ depends on the choice of the gauge-fixing condition $G$. 
Combining these results together we find
\begin{equation}
Z_{\rm YM} = \int \!{\cal D}{\cal B}_n^\mu \, \, \delta[\bn \mcdot {\cal B}_n]\,J_G[{\cal B}_n]\,E_G[A_n({\cal B}_n)]\,.
\end{equation}
Everything in this generating functional is known, except for the explicit form of the Jacobian $J_G[{\cal B}_n]$ and the determinant inside $E_G[A_n(\cB_n)]$. Due to the non-linear nature of Eq.~(\ref{ABrelation}), their expressions for a general gauge-fixing condition $G$ are very difficult to derive. However, from the Faddeev-Popov procedure it is obvious that the generating functional is identical for all choices of the gauge-fixing condition $G$, since it was introduced as an arbitrary function in 
Eq.~(\ref{FPunity}). Thus, any choice of $G(A_n)$ will do, and the easiest choice is light-cone gauge, which uses 
\begin{equation}
G(A_n) \equiv G_{\rm LC}(A_n) = \bn \mcdot A_n \,.
\end{equation}
In this case, we have $\bn \mcdot A_n = 0$, which immediately implies $W_n = 1$, making the relation between the $A_n^{\mu}$ and ${\cal B}_n^{\mu}$ fields trivial:
\begin{equation}
{\cal B}_n^\mu = A_n^\mu\,.
\end{equation}
Thus, in this particular gauge we find
\begin{equation}
J_{G_{\rm LC}}[{\cal B}_n] = 1 \,, \qquad E_{G_{\rm LC}}[{\cal B}_n] = {\mathrm{det}}(\bn \mcdot  \partial) \,e^{iS_{\rm YM}[{\cal B}_n]}\,,
\end{equation}
and we obtain the final form of the generating fuctional in terms of $\cB^{\mu}_n$ fields as
\begin{equation}\label{Bactionfinal}
Z_{\rm YM} = \int \!{\cal D}{\cal B}_n^\mu \, \delta[\bn \mcdot {\cal B}_n]\,{\mathrm{det}}(\bn \mcdot \partial) \,e^{iS_{\rm YM}[{\cal B}_n]}\,.
\end{equation}
In other words, the Yang-Mills action in terms of the gauge invariant gluon field ${\cal B}_n^\mu$ is identical to the one in terms of the field $A_n^\mu$ in the light-cone gauge. Therefore, all Feynman rules for the ${\cal{B}}_n^{\mu}$ fields are identical to Feynman rules for the $A_n^{\mu}$ fields in the light-cone gauge. In particular, this justifies Eq.~(\ref{propB}) as the right form of the gluon propagator for the $\cB_n^{\mu}$ fields. Incidentally, notice also that the determinant in Eq.~(\ref{Bactionfinal}) is independent of the gauge field and therefore can be ignored, meaning that the formulation with $\cB_n^{\mu}$ fields is ghost-free. This obviously complies with the well known fact that the light-cone gauge is unitary and ghost fields decouple (see, for instance, \cite{tHooft}). 

Having worked out how one can quantize SCET directly in terms of the gauge invariant degrees of freedom, we next ask whether it is possible to write the generating functional of SCET in terms of fields, whose interactions are given by the interactions of full QCD. 
As we will show, this is indeed possible if we restrict ourselves to leading order in the power counting, but requires separate fields for each different collinear direction. We will also show how to construct external operators coupling these different fields to one another, such that any leading order correlation function in SCET can be reproduced using only fields whose coupling to other fields is described by the Lagrangian of full QCD. We do want to emphasize that {\em this by no means implies} that SCET as an effective theory is useless. The power of SCET comes from understanding the interactions between fields in different directions, and while we can reproduce any leading order operator using fields that resemble full QCD, we can neither easily implement power corrections, nor can we derive the form of the leading order operators without the construction of SCET. However, we can use this equivalence to calculate matrix elements in SCET using the familiar Feynman rules of QCD, which will in general simplify the required calculations at higher orders in perturbation theory. 

We start by making the Ansatz
\begin{equation}
\label{Ansatz}
Z[J]=\int \!{\cal{D}}{\bar{\psi_n}}{\cal{D}}\psi_n {\cal{D}}A_n^{\mu} {\mathrm{exp}}\left[i\int  \! d^4x\, S_{\rm QCD}(\psi_n,A_n,J) \right],
\end{equation} 
where $S_{\rm QCD}$ is defined by
\begin{eqnarray}
\label{QCDaction}
S_{\rm QCD} &\!\!=\!\!& \sum_n\big[{\cal{L}}^{\rm QCD}_n+{\bar{J}}^\xi_n {\cal{M}}^\xi_n\psi_n+{\bar{\psi}}_n{\bar{{\cal{M}}}}^\xi_nJ^\xi_n+{\bar{J}}^\chi_n {\cal{M}}^\chi_n\psi_n\nn\\
&&+{\bar{\psi}}_n{\bar{{\cal{M}}}}^\chi_nJ^\chi_n+ J_{n\mu}^A A_n^\mu+J_{n\mu}^{\cal B}{\cal B}_n^\mu(A_n)\big]\nn\\
&&+\sum_{k}J_{k} {\cal{Q}}_{k}(\psi_n,A_n)
\,,\end{eqnarray}
with 
\begin{equation}
{\cal L}_n^{\rm QCD} = \bar \psi_n\, i \Dslash \, \psi_n\,.
\end{equation}
The set of operators ${\cal Q}_k$ couple $k$ fields in different directions $n_1,...,n_k$.
Our goal is to find expressions for ${\cal M}_n$ and ${\cal Q}_k$, such that the generating functional in Eq.~(\ref{Ansatz}) is equivalent to the generating functional of SCET.

Let's begin by setting all currents in the action to zero, leaving only the Lagrangian ${\cal L}_n^{\rm QCD}$. One can write
\begin{equation}
\psi_{n}(x)=\left(P_n+P_\bn\right)\psi_{n}(x)\,,
\end{equation}
with the projection operators $P_n$ and $P_\bn$ defined by
\begin{equation}
\label{Pndef}
P_n = \frac{\nslash\bnslash}{4}\,, \qquad P_\bn = \frac{\bnslash\nslash}{4}\,,
\end{equation}
and define
\begin{equation}
\xi_{n} \equiv P_n\psi_{n}\,,\qquad \phi_{n} \equiv P_\bn\psi_{n}\,.
\end{equation}
This allows us to write
\begin{eqnarray}
Z[J=0]&=& \int {\cal{D}}{\bar{\xi}_n}{\cal{D}}\xi_n {\cal{D}}{\bar{\phi}_n}{\cal{D}}\phi_n {\cal{D}}A_n^{\mu}\\
&& \times {\mathrm{exp}}\left[\sum_n i \!\int  \! d^4x\,(\bar \xi_n + \bar \phi_n) i\Dslash \, (\xi_n + \phi_n) \right]\,.\nn
\end{eqnarray} 
Using the well-known formula for Gaussian integration,
\begin{eqnarray}
&&\int \!{\cal{D}}\phi{\cal{D}}{\bar{\phi}}\,\, {\mathrm{exp}}\left[i\int \!d^4x ({\bar{\phi}}M\phi + {\bar{J}}\phi + {\bar{\phi}} J)\right]\nn\\
&& \qquad ={\mathrm{det}}(-i M)\,{\mathrm{exp}}\left[-i\int \!d^4x {\bar{J}}\frac{1}{M} J\right]\,,\label{MasterFormula}
\end{eqnarray}
it is straightforward to perform the integrals over $\phi_n$ explicitly. We find
\begin{equation}
Z[J=0]= \int \!{\cal{D}}{\bar{\xi}_n}{\cal{D}}\xi_n  {\cal{D}}A_n^{\mu} {\mathrm{exp}}\left[i \!\int  \! d^4x\, \sum_n{\cal L}_n^{\rm SCET}\right]\,,\label{ZJ0}
\end{equation} 
where
\begin{equation}
\label{LSCET}
{\cal L}_n^{\rm SCET} = {\bar{\xi}}_n\left[i n\mcdot D+i\Dslash_{\perp}\frac{1}{i{\bar{n}}\mcdot D}i\Dslash_{\perp}\right]\frac{\bnslash}{2}\xi_n\,.
\end{equation}
Note that in getting to Eqs.~(\ref{ZJ0}) and (\ref{LSCET}) no expansion has been made, only integration of modes in the generating functional. Also note that in Eq.(\ref{ZJ0}) we have omitted the determinant factor in Eq. (\ref{MasterFormula}). Indeed it is easy to show that 
\begin{align}
&\det\left(\frac{\nslash}{2}\bar{n}\mcdot D\right)=\int \!{\cal{D}}\eta_n{\cal{D}}{\bar{\eta}_n}\,\, {\mathrm{exp}}\left[-\!\int \!d^4x \, {\bar{\eta}_n}\left(\frac{\nslash}{2}\bar{n}\mcdot D\right)\eta_n\right]\nonumber\\
&=\int \!{\cal{D}}\eta_n^{\prime}{\cal{D}}{\bar{\eta}}_n^{\prime}\,\, {\mathrm{exp}}\left[-\!\int \!d^4x \,{\bar{\eta}}_n^{\prime}\left(\frac{\nslash}{2}W^{\dagger}_n\bar{n}\mcdot D W_n\right)\eta_n^{\prime}\right]\nonumber\\
&=\det\left(\frac{\nslash}{2}\bar{n}\mcdot \partial\right),
\end{align}
where we have defined $\eta_n^{\prime}=W_n^{\dagger} \eta_n$. Thus the determinant is just an overall constant and can be ignored.
 
We can now move on and consider the addition of current terms in the action. Keeping the currents ${\bar J}_n {\cal M}_n$ and ${\bar {\cal M}}_nJ_n$ for the fields $\psi_n$ and $\bar \psi_n$, but still neglecting the currents $J_k$ for the local operators ${\cal Q}_k$, and again performing the integrals over $\phi_n$ and $\bar \phi_n$ gives
\begin{eqnarray}
Z[J_k=0]= \int \! {\cal{D}}{\bar{\xi}_n}{\cal{D}}\xi_n  {\cal{D}}A_n^{\mu}\,
{\mathrm{exp}}\left[i \!\int  \! d^4x\, S^{\rm SCET}(J_k=0)\right]\,,\nonumber\\
&&
\end{eqnarray} 
with
\begin{widetext}
\begin{eqnarray}
\label{actionJkis0}
S^{\rm SCET}(J_k=0)&=&\sum_n {\cal L}_I^n +{\bar{J}}^\xi_n{\cal{M}}^\xi_n{\cal{R}}_n \xi_n+{\bar{\xi}}_n{\bar{\cal{R}}}_n{\bar{\cal{M}}}^\xi_n J^\xi_n
+ {\bar{J}}^\chi_n{\cal{M}}^\chi_n{\cal{R}}_n \xi_n+{\bar{\xi}}_n{\bar{\cal{R}}}_n{\bar{\cal{M}}}^\chi_n J^\chi_n+J_{n\mu}^A A_n^\mu+J_{n\mu}^{\cal B}{\cal B}_n^\mu(A_n)
\nonumber\\
&& 
- \left(\bar J_n^\xi {\cal M}_n^\xi +  \bar J_n^\chi {\cal M}_n^\chi  \right)\frac{1}{i \bn \mcdot D} \frac{\bnslash}{2} \left(\bar{\cal M}_n^\xi J_n^\xi + \bar{\cal M}_n^\chi J_n^\chi  \right)\,.
\end{eqnarray}
\end{widetext}
Here we have defined
\begin{equation}
{\cal{R}}_n=\left[1+\frac{1}{i{\bar{n}}\mcdot D}i\Dslash_{\perp}\frac{\bnslash}{2}\right]\,.
\end{equation}

In order for this action to be equal to the action of SCET given in Eq.~(\ref{actionSI}) (still with $J_k=0$), requires 
\begin{eqnarray}
\label{Msols}
{\cal{M}}^\xi_n{\cal{R}}_n \xi_n \equiv \xi_n\,, \qquad
{\cal{M}}^\chi_n{\cal{R}}_n \chi_n \equiv W_n^\dagger\xi_n\,,
\end{eqnarray}
in addition to having the second line in Eq.~(\ref{actionJkis0}), corresponding to contact terms arising when taking two derivatives of the generating functional with respect to the currents $J_n^{\xi/\chi}$, vanish. 
There are two possible solutions for each of the ${\cal{M}}^\xi_n$ and ${\cal{M}}^\chi_n$ to satisfy Eq.~(\ref{Msols}), namely
\begin{align}
\label{SnJkis0}
{\cal{M}}^\xi_n&={\cal{R}}_n^{-1}&{\rm or} \qquad
{\cal{M}}^\xi_n&=P_n\,,
\nonumber\\
{\cal{M}}^\chi_n&=W_n^{\dagger}{\cal{R}}_n^{-1}& {\rm or} \qquad
{\cal{M}}^\chi_n&=W_n^{\dagger}P_n\,.
\end{align}
While both of these solutions for ${\cal M}_n$ give the same answer, the second choice is in practice much easier to use.  This is because choosing ${\cal{M}}^\xi_n= {\cal R}_n^{-1}$ in Eq.~(\ref{QCDaction}) adds couplings between fermions and gluons to the current terms, complicating perturbative calculations significantly. Furthermore, for the second solution the second line in Eq.~(\ref{actionJkis0}) vanishes as desired. Therefore, for ${\cal{M}}^\xi_n=P_n$ and ${\cal{M}}^\chi_n=W^{\dagger}_nP_n$ we obtain for $J_k=0$ the desired result $S^{\rm SCET} = S_I$, where $S_I$ is defined in Eq.~(\ref{actionSI}).

Finally, we add the currents for the local operators ${\cal Q}_k$ back to the action. Since these operators couple fields with different $n$'s to one another, integrating out the $\phi_n$ fields is very complicated. However, there is a simple choice for the operators ${\cal Q}_k$ that will directly reproduce the form $\sum_k J_k {\cal O}_k$ present in the final answer, Eq.~(\ref{actionSI}). This is achieved by taking
\begin{equation}
{\cal Q}_k(\psi_n,A_n) = {\cal O}_k(W_n^\dagger P_n \psi_n,\cB_n(A_n))\,,
\end{equation}
with $P_n$ defined in Eq.~(\ref{Pndef}). Since $P_n \psi_n = \xi_n$, this choice eliminates any dependence on $\phi_n$ in ${\cal Q}_k$. Thus, the integrals over $\phi_n$ can be performed as before and we therefore find 
\begin{equation}
{\cal Q}_k(\psi_n,A_n) = {\cal O}_k(W_n^\dagger \xi_n,\cB_n(A_n))\,.
\end{equation}

In conclusion, the generating functional in terms of QCD fields
\begin{equation}
Z[J]=\int {\cal{D}}{\bar{\psi_n}}{\cal{D}}\psi_n {\cal{D}}A_n^{\mu} {\mathrm{exp}}\left[i\int  \! d^4x\, S_{\rm QCD}(\psi_n,A_n,J) \right]\,,
\end{equation} 
with $S_{\rm QCD}$ defined by
\begin{eqnarray}\label{answer}
S_{\rm QCD} &\!\!=\!\!& \sum_n\big[
{\cal{L}}^{\rm QCD}_n
+{\bar{J}}^\xi_n P_n\psi_n
+{\bar{\psi}}_n P_\bn J^\xi_n
+{\bar{J}}^\chi_n W_n^\dagger P_n\psi_n
\nn\\
&&
+{\bar{\psi}}_n P_\bn W_n J^\chi_n
+ J_{n\mu}^A A_n^\mu+J_{n\mu}^{\cal B}{\cal B}_n^\mu(A_n) \big]
\nn\\
&&
+\sum_{k}J_{k} {\cal{O}}_{k}(W^\dagger_n P_n \psi_n,\cB_n(A_n))
\,,\end{eqnarray}
is identical to the generating functional defined in Eqs.~(\ref{Zdef}) and~(\ref{actionSI}) in terms of SCET fields. This proves that the collinear sector of SCET is equivalent to a theory containing multiple copies of QCD, where the only interactions between them are contained in the local operators ${\cal O}_k$. 

So far we have only considered the collinear sector of SCET, but of course it is well known that usoft degrees of freedom are required in order to reproduce the long distance dynamics of QCD. On the other hand, it is also well known that at leading order in the effective theory the interactions between usoft and collinear particles can be removed to all orders in perturbation theory by using the field redefinition \cite{Bauer:2001yt}
\begin{equation}
\xi_n \to Y_n \xi_n\,,
\end{equation}
where
\begin{equation}
Y_n = {\rm P} \exp \left[ i g \! \int_0^\infty \! {\rm d} s \,\,n \mcdot A(n s + x) \right]\,.
\end{equation}
Thus, we can include the interactions with the usoft gluons by making a similar field redefinition on the fields $\psi_n$. This implies that the action given in Eq.~(\ref{QCDaction}), but now with 
\begin{equation}
{\cal L}_n^{\rm QCD} = \bar \psi_n\, \left( i \Dslash + g n \mcdot A_{us} \frac{\bnslash}{2} \right)\, \psi_n\,
\end{equation}
reproduces both the collinear and usoft interactions of the collinear fields.

In conclusion, we have shown how SCET can be quantized either in terms of gauge dependent or gauge invariant fields. In practice, most calculations in the literature are performed using the gauge dependent degrees of freedom, whereas the external operators have to depend on the gauge invariant fields. Using our results, one can perform the calculations directly in terms of the gauge invariant fields, reducing the number of Feynman diagrams significantly. We have then moved on to show how the collinear sector of SCET is equivalent to a theory constructed out of multiple decoupled copies of full QCD, in the sense that each copy describes the interactions of fields in a given direction and the different copies do not interact with one another. We have also shown in detail how to construct the local operators describing precisely the interactions between the different copies of QCD, such that any SCET correlator at leading order can be reproduced.

\begin{acknowledgments}
We would like to thank Iain Stewart for helpful discussions. %
This work was supported in part by the Director, Office of
Science, Office of High Energy Physics of the U.S.\ Department of Energy under
the Contract DE-AC02-05CH11231.
CWB acknowledges support from an LDRD grant from LBNL.
OC would like to thank the Fulbright Program and the Spanish Ministry of Education and Science for financial support under grant number FU2005-0791 and also LBNL and UCB-BCTP for their hospitality this past year.
\end{acknowledgments}


\begin{appendix}

\section*{Appendix: Diagrammatic proof of the equivalence of QCD and SCET with one collinear direction}

In the main body of this paper we have shown that any 
collinear SCET diagram can be obtained using a gener- 
ating functional in which the interactions between the fields are equivalent to full QCD, but the external cur- 
rents are modified to contain projection operators. This 
relation was first  discussed in~\cite{Bauer:2000yr} and used in~\cite{Becher:2006qw} 
to calculate jet functions in SCET. In this appendix we want 
to prove this identity diagrammatically for the correlator 
containing two collinear fermions and $N$ collinear gluons. 

We will accomplish this by working out in both theories the Feynman 
diagrams for $N$ gluons coupled to a fermion line, from which the correlator can be constructed. 
Using this result we will then show that both of these calculations lead to equivalent answers. 
Note that there are $N!$ possible color structures, and for 
each of them the QCD result has to equal 
the SCET result. We begin by showing this equivalence 
for the color structure $T^{a_1}\, T^{a_2}\, \ldots \, T^{a_N}$, 
and then discuss 
how the result can be modified to include the other color 
structures as well.

Define $Q^{(N)}$ and $S^{(N)}$ to be the QCD and SCET cor- 
relators for this color structure in momentum space, multiplied 
by a factor of $p_i^2$ for each internal propagator and with 
the factor $g_s^N$ removed. This gives
\begin{align}
&Q^{(N)}=P_n \dslash{p}_0 \gamma^{\mu_1} \dslash{p}_1 \dots  \gamma^{\mu_N} \dslash{p}_N P_{\bar{n}}
\label{Qdef}\, ,\\
&S^{(N)}=\sum_{k=1}^N S^{(N-k)} L_k \,.
\label{Sdef}
\end{align}
The first equation follows simply from the QCD Feynman rules, while the SCET equation is a recurrence formula, that takes into account all the possibilities of having $k$ out of the $N$ gluons being emitted from a single vertex. $L_k$ is therefore the Feynman rule for $k$-gluon emissions from a single vertex, multiplied by a factor of $\prod_i p_i^2/g_s$ to account for the removal of the factors $p_i^2$ and $g_s$, as discussed above:
\begin{equation}
L_k \frac{\bnslash}{2} = i (\bar{n}\mcdot p_N)\frac{p_{N-k+1}^2 \cdots p_{N-1}^2}{(-g)^k} V_k\,,
\end{equation}
with
\begin{align}
&V_k = \frac{i(-g)^k \bar{n}^{\mu_{N-k+2}} \dots\bar{n}^{\mu_{N-1}}}{\bar{n}\mcdot p_{N-k+1}\dots \bar{n}\mcdot p_{N-1}}  \frac{\dslash{\bar{n}}}{2}\times\\
&\times\Big(\gamma^{\mu_{N-k+1}}_\perp \gamma^{\mu_N}_\perp-\bar{n}^{\mu_{N}}\gamma^{\mu_{N-k+1}}_\perp \frac{\dslash{p}_{N}^\perp}{\bar{n}\mcdot p_N}+\nonumber\\
&+\frac{\dslash{p}_{N-k}^\perp \dslash{p}_{N}^\perp}{\bar{n}\mcdot p_{N-k} \bar{n}\mcdot p_{N}}\bar{n}^{\mu_{N-k+1}}\bar{n}^{\mu_{N}}-\bar{n}^{\mu_{N-k+1}}\frac{\dslash{p}_{N-k}^\perp\gamma^{\mu_{N}}_\perp}{\bar{n}\mcdot p_{N-k}}\Big)\,.\nn
\end{align} 

We will show the equivalence $Q^{(N)} = S^{(N)}$ by induction. For $N=0$ it is straightforward:
\begin{align}
Q^{(0)}&=P_n \dslash{p}_0 P_{\bar{n}}=\frac{\dslash{n}}{2} \bar {n} \mcdot p_0, \\
S^{(0)}&=\frac{\dslash{n}}{2} \bar {n} p_0 = Q^{(0)}\,.
\end{align}
Next, we assume that the statement $Q=S$ holds for $0,1,\dots N-1$ to show that this leads to $Q^{(N)}=S^{(N)}$. This implies
\begin{equation}
Q^{(N)}=\sum_{k=1}^N Q^{(N-k)} L_k  \,.
\label{maineq}
\end{equation}
To prove Eq.~(\ref{maineq}) we rewrite the general QCD correlator $Q^{(N)}$ by pushing the projection operator $P_n$ in Eq.~(\ref{Qdef}) through the $\dslash{p}_n$ and $\gamma^{\mu_n}$, to obtain
\begin{equation}
Q^{(N)}=\frac{\dslash{n}}{2}\sum_{m=0}^N\sum_{l=1}^{C_{2N+1}^{2m}} \chi^{2m, 2N+1}_{i_1\dots i_{2m}}\,,
\label{Qexpan}
\end{equation}   
where
\begin{align}
&\chi^{2m, 2N+1}_{i_1\dots i_{2m}}={(-1)}^{i_1+\dots +i_{2m}-(1+\dots+2m)}\perp_{i_1}\dots \perp_{i_{2m}}
\nonumber\\
  &\quad\quad\quad\times(\bar{n}_{j_1} n_{j_2} \bar{n}_{j_3}\dots n_{j_{2N-2m}} \bar{n}_{j_{2N+1-2m}})\,.
\end{align}
Here $C_{k}^{l}$ denotes the binomial coefficient for $l$ choose $k$, and we have used a shorthand notation in which $\bar{n}_{j}$ corresponds to $\bar{n}\mcdot p$ for even $j$ and to $\bar{n}^{\mu}$ for odd $j$ and accordingly $\perp_{j}$ corresponds to $\dslash{p}_\perp$ for even $j$,  while $\gamma_\perp^\mu$ for odd $j$. 

We would like to comment on how we obtained this result. Expanding each $\gamma$ matrix on the right hand side of the Eq. (\ref{Qdef}) according to $\gamma^{\alpha}=\bar{n}^{\alpha}\frac{\dslash{n}}{2}+n^{\alpha}\frac{\dslash{\bar{n}}}{2}+\gamma^{\alpha}_\perp$ will result in terms with fixed number $0 \le N_\perp \le 2N+1$ of $\gamma_\perp$'s, together with $(2N+1-N_\perp)$ of $\dslash{n}$ or $\dslash{\bn}$. Since the $\dslash{n}$ and $\dslash{\bn}$ terms have to alternate, and the projection operator forces the first and last term to be $\dslash{n}$, $N_\perp$ has to be an even number. 

As a next step, we work out the sum on the right hand side of Eq.~(\ref{maineq}).  Note that the term $L_k$ contains factors of $p_i^2$ in the numerators, while there are no such terms on the left hand side of  Eq.~(\ref{maineq}). However, both $Q^{(N-k)}$ and $L_k$ contain terms with $\dslash{p}_{\perp}$, which can lead to $p_{\perp}^2 = p^2 - n \mcdot p \, \bn \mcdot p$. After a straightforward, but lengthy calculation, one can show that
\begin{equation}
\sum_{k=1}^N Q^{(N-k)} L_k =\frac{\dslash{n}}{2}\sum_{m=0}^{N}\sum_{l=1}^{C_{2N+1}^{2m}} \chi^{2m, 2N+1}_{i_1\dots i_{2m}}\,.
\end{equation}   
Thus, both sides of Eq.~(\ref{maineq}) are equal and we have thus shown that $Q^{(N)} = S^{(N)}$ for all values of $N$.

So far we have only dealt with the term with color structure $T^{a_1}\dots T^{a_N}$. Keeping the general color structure allows us to write
\begin{align}
&Q^{(N)}\rightarrow \sum_{l=1}^{N!} Q^{(N)}_{i_1\dots i_N} T^{a_{i_1}}\dots T^{a_{i_N}}\,,\\
&S^{(N)}\rightarrow \sum_{l=1}^{N!} S^{(N)}_{i_1\dots i_N} T^{a_{i_1}}\dots T^{a_{i_N}}\,.
\end{align}
What we have shown so far is that $Q^{(N)}_{1,2,\dots, N} =S^{(N)}_{1,2,\dots, N}$. However, it is clear that the proof goes through for any color permutation, with obvious replacements to account for the different orderings of the gluons. Finally, notice that triple or quartic gluon vertices do not change the result, since they are the same in QCD and SCET. This completes the proof.

\end{appendix}

\end{document}